\documentclass[12pt]{article}
\usepackage{scicite}
\usepackage{times}
\newenvironment{sciabstract}{%
\begin{quote} \bf}
{\end{quote}}

\newcounter{lastnote}
\newenvironment{scilastnote}{%
\setcounter{lastnote}{\value{enumiv}}%
\addtocounter{lastnote}{+1}%
\begin{list}%
{\arabic{lastnote}.}
{\setlength{\leftmargin}{.22in}}
{\setlength{\labelsep}{.5em}}}
{\end{list}}

\usepackage[dvips,usenames]{color}
\usepackage{txfonts}
\usepackage{graphicx}

\title{Universality in Three- and Four-Body Bound States of Ultracold Atoms}

\author
{Scott E. Pollack,$^{\ast}$ Daniel Dries, Randall G. Hulet\\
\\
\normalsize{Department of Physics and Astronomy and Rice Quantum Institute,}\\
\normalsize{Rice University, Houston, TX  77005, USA (\small November 4, 2009)}\\
\\
\normalsize{$^\ast$To whom correspondence should be addressed; E-mail:  scott.pollack@rice.edu}
}

\date{}

\begin{document} 

\maketitle 
\baselineskip20pt

\begin{sciabstract}
Under certain circumstances, three or more interacting particles may form bound states.  
While the general few-body problem is not analytically solvable, 
the so-called Efimov trimers appear for a system of three particles 
with resonant two-body interactions.
The binding energies of these trimers are predicted to be universally connected to each other, 
independent of the microscopic details of the interaction.
By exploiting a Feshbach resonance to widely tune 
the interactions between trapped ultracold lithium atoms, 
we find evidence for two universally connected Efimov trimers
and their associated four-body bound states.
A total of eleven precisely determined three- and four-body features 
are found in the inelastic loss spectrum.
Their relative locations on either side of the 
resonance agree well with universal theory,
while a systematic deviation from universality is found when
comparing features across the resonance.
\end{sciabstract}

\clearpage
\baselineskip15pt


One of the most remarkable few-body phenomena is the universally-connected 
series of three-body bounds states first predicted by Efimov \cite{efimov71}.  
Efimov showed that three particles can bind in the presence of 
resonant two-body interactions, even in circumstances where 
any two of the particles are unable to bind.  
When the two-body scattering length $a$ is much larger than the range of the 
interaction potential $r_0$, the three-body physics becomes independent of the 
details of the short-range interaction.
Surprisingly, if one three-body bound state exists, then another can be found by 
increasing $a$ by a universal scaling factor, and so on, resulting in an infinite 
number of trimer states \cite{braaten_physrep}.  
Universality is expected to persist with the addition of a 
fourth particle \cite{four1,four3,hammer_platter,wang:133201,javier}, 
with two four-body states associated with each trimer \cite{hammer_platter,javier}; 
intimately tied to the
three-body state, these tetramers do not require any additional parameters
to describe their properties.

Ultracold atoms are ideal systems for exploring these weakly bound few-body states
because of their inherent sensitivity to low-energy phenomena, 
as well as the ability afforded by Feshbach resonances to continuously 
tune the interatomic interactions.  
Pioneering experiments with trapped, ultracold atoms have obtained signatures of individual 
Efimov states \cite{Kraemer,knoop,ottenstein:203202,ohara,barontini:043201},
as well as two successive Efimov states \cite{Zaccanti,Khaykovich},
via their effect on inelastic collisions that lead to trap loss.  
Evidence of tetramer states associated with the trimers 
has also been found \cite{ferlaino:140401,Zaccanti}.
Although the locations of successive features are consistent with the 
predicted universal scaling, systematic deviations as large as 60\% were observed, 
and attributed to non-universal short-range physics \cite{Zaccanti}.  
In the work presented here,
we use a Feshbach resonance in $^7$Li for which $a/r_0$ can be tuned 
over a range spanning 3 decades \cite{Pollack}.
This enables the observation of multiple features which are compared to universal theory.  


We confine $^7$Li in the $|F = 1, m_F = 1\rangle$ hyperfine state in
an elongated, cylindrically symmetric, hybrid magnetic plus optical
dipole trap, as described previously \cite{Pollack}.
A set of Helmholtz coils provides an axially oriented magnetic bias field
used to tune the two-body scattering length $a$ via a Feshbach resonance
located near $737\,G$ \cite{som}.  For $a > 0$, efficient evaporative
cooling is achieved by setting the bias field to 717\,G, 
where $a \sim 200\,a_0$ (with $a_0$ the Bohr radius), and reducing the optical trap intensity.  
Depending on the final trap depth, we create either an ultracold thermal gas just above the
condensation temperature $T_C$, or a Bose-Einstein condensate (BEC) with $>90\%$
condensate fraction.  For investigations with $a < 0$, we first set the
field to 762\,G where $a \sim -200\,a_0$ and proceed with optical trap evaporation, 
which is stopped at a temperature $T$ slightly above $T_C$.
In both cases the field is then adiabatically
ramped to a final value of $a$ and held for a variable hold time.
The fraction of atoms remaining at each time is measured 
via \emph{in situ} polarization phase contrast imaging \cite{PhysRevLett.78.985}
for clouds 
where the density is high, or absorption imaging in the case of lower densities.

Analyzing the time evolution of the number of atoms in the trap determines
the three-body loss coefficient $L_3$ \cite{weber,Kraemer,Zaccanti}
as well as the four-body loss coefficient $L_4$ \cite{ferlaino:140401}.  
Recombination into a dimer is a three-body process since a third atom 
is needed to conserve both momentum and energy.  
For $a > 0$, the dimer can be weakly-bound with binding energy 
$\epsilon = \hbar^2/(m a^2)$, where $m$ is the atomic mass, 
while for $a < 0$ there are only deeply-bound molecular dimers.  
The recombination energy released in the collision is sufficient 
to eject all three atoms from the trap for $a < 0$, and for $a > 0$ when $\epsilon \gtrsim U$, 
where $U$ is the trap depth.  
In the case of the BEC data, this latter condition holds for $a \lesssim 5000\,a_0$.  
Nonetheless, we assume that all three atoms are lost for any recombination event,
because even for $a$ larger than $5000\,a_0$ we observe rapid three-body loss.  
We ascribe this observation to a high probability for dimers to undergo 
vibrational relaxation collisions which result in kinetic energies much greater than $U$.  
Four-body processes proceed in a similar fashion \cite{wang:133201,ferlaino:140401}.

The equation describing the dynamics of three- and four-body loss is
\begin{equation}
\frac{1}{N}\frac{dN}{dt} = - \frac{g^{(3)}}{3!} L_3 \langle n^2 \rangle - \frac{g^{(4)}}{4!} L_4 \langle n^3 \rangle,
\end{equation}
where the brackets denote averages over the density distribution \cite{som}.
For a thermal gas the spatial correlation 
coefficients $g^{(3)}$ and $g^{(4)}$ are 
respectively $3!$ and $4!$, while for a BEC we set both to 1\cite{kagan,PRL.79.337}.  
We have verified that heating from
recombination is small for our short observation times
and therefore omit this effect in our analysis \cite{weber,ferlaino:140401}.
By fitting the time evolution of the number of atoms to the solution
of Eq.~1 we extract $L_3$ and $L_4$ as a function of~$a$.
Figure~S1 shows the loss of atoms as a function of time in regimes where
either $L_3$ or $L_4$ dominates \cite{som}.
Four-body loss is readily distinguished from three-body loss by
the shape of the loss curve.

Figure~1 shows the extracted values of $L_3$ across the Feshbach resonance,
exhibiting the expected $a^4$ scaling \cite{NielsenMacek,PhysRevLett.83.1751},
but with several dips and peaks punctuating this trend.  
Two prominent peaks dominate the landscape for $a < 0$, which are 
labeled $a^-_1$ and $a^-_2$ in Fig.~1A.  
We attribute these peaks to the crossings of the energies of the first two
trimer states with the free atom threshold,
thus providing additional pathways into deeply-bound molecular states \cite{PhysRevLett.83.1751}. 
For $a > 0$, the dominant features are dips, indicated in Fig.~1A as $a^+_1$ and $a^+_2$, 
corresponding to recombination minima.
These minima are associated with the merging of the same two trimer 
states into the atom-dimer continuum, and have been attributed to 
destructive interference between 
two different decay pathways into weakly-bound dimers \cite{NielsenMacek,PhysRevLett.83.1751}.
We fit the data to $L_3(a) = 3 C(a) \hbar a^4 / m$, 
where $C(a)$ is a logarithmically periodic function characterizing
effects from the Efimov states \cite{som}.  The analytic expression for $C(a)$
contains the location of one universal trimer resonance $a^{-}\!<0$ or
recombination minimum $a^+ > 0$, and an inelasticity parameter $\eta$
related to the lifetime of the Efimov state \cite{braaten_physrep}.  
The observed features are fit individually to extract these parameters (Table~1).
The universal theory describing
Efimov physics \cite{braaten_physrep} predicts a logarithmic spacing in
the two-body scattering length between trimer states of $e^{\pi/s_0} \approx 22.7$,
where $s_0 = 1.00624$ is a universal parameter \cite{efimov71}.
Table~2 shows that the ratios $a^+_2/a^+_1$ and $a^-_2/a^-_1$ 
agree well with the universal theory.

A local maximum in $L_3$, indicated as $a^*_2$ and shown in detail in Fig.~1B,
can be discerned between the two recombination minima $a^+_1$ and $a^+_2$.
We associate this feature with an atom-dimer resonance,
given its location with respect to the nearby minima.  
A simple model \cite{Zaccanti} has been proposed to explain 
the enhanced losses present at the atom-dimer resonance.  This model
describes an avalanche process whereby a single dimer travelling through
a collisionally thick gas shares its kinetic energy with multiple atoms,
thereby increasing from 3 the effective number of atoms 
lost for each dimer formed \cite{Schuster}.  

For $a < 0$, $L_3$ achieves its maximum value of $\sim$$10^{-19}\,\mathrm{cm^6/s}$
at $a^-_2$.  This value is reasonably consistent with the 
expected unitarity limit \cite{weber,dincao:123201}.
At even larger values of $|a|$, $L_3$ saturates to a value below the unitarity limit, 
a behavior previously seen in experiment \cite{Kraemer} and in numerical
calculations \cite{dincao:123201,dincao_jpb}.


The four-body loss coefficient $L_4$ for $a < 0$ was also extracted from the data, 
and the results are presented in Fig.~2.  
Three resonant peaks in $L_4$ are observed, 
which we associate with the crossings of tetramer states with the free atom 
continuum \cite{four1,hammer_platter,four3,wang:133201,javier,ferlaino:140401,Zaccanti,mehta}.  
Two universal tetramers are predicted to accompany each 
Efimov trimer \cite{hammer_platter,javier}.  
The solid line in Fig.~2 is calculated using only the 
observed three-body locations and widths in addition to an overall scaling, 
without any other free parameters \cite{som}.  
The agreement between this curve and the data lead us to assign the peaks 
to the second tetramer of the first Efimov trimer $a^T_{1,2}$ and 
both tetramers of the second Efimov trimer $a^T_{2,1}$ and $a^T_{2,2}$ \cite{ferlaino:140401}.  
While we do not have the resolution to detect an enhancement in $L_4$ at 
the expected location of the first tetramer $a^T_{1,1}$, 
an enhancement of $L_3$ is observed at the expected location (Fig.~1A) which we 
tentatively identify with $a^T_{1,1}$ \cite{javier,Zaccanti}. 
The existence of two tetramer states tied to a single trimer state
has also been verified in $^{133}$Cs \cite{ferlaino:140401} and $^{39}$K \cite{Zaccanti}.  

Two additional peaks in $L_3$ are observed on the $a > 0$ 
side of the resonance (Figs.~1C and 1D).
Features at these relative positions have not been previously observed or predicted, 
although they occur very close to where the two tetramer states associated with the
second trimer are expected to merge with the dimer-dimer continuum \cite{dincao:033004}.
We have no explanation of how a dimer-dimer resonance would affect the inelastic loss
rate, as we expect the dimer fraction to be small and consequently, 
the probability of dimer-dimer collisions to be negligible.
One possibility is that they arise because of an interference effect,
similar to that occurring in the three-body process at $a^+_1$ and $a^+_2$.
Presently, we tentatively associate these features with 
dimer-dimer resonances located at $a^*_{2,1}$ and $a^*_{2,2}$.

In Table~2 we present the relative spacings of observed loss features
along with those predicted by the universal theory.  
Universal scaling is expected when $|a| \gg r_0$, 
where $r_0$ is the van der Waals radius ($33\,a_0$ for Li) \cite{kohler:1311}.  
Another requirement for universality is that $|a| \gg |R_e|$, 
where $R_e$ is the effective range \cite{Khaykovich}.
Figure~S4 shows that $R_e$ is relatively small over the relevant field range,
and is $\sim$$-10\,a_0$ on resonance \cite{som}.   
For comparison, in the $|1,0\rangle$ state of $^7$Li,
$R_e \sim -30\,a_0$ at the resonance near 894\,G \cite{Khaykovich}.
Both conditions for universality are well-satisfied for the second Efimov state, 
but the requirement that $|a| \gg r_0$ is only marginally satisfied for the first.  
Nonetheless, we find good agreement with the universal scaling relations 
between features on each side of the Feshbach resonance separately.

The relationships of features across a Feshbach resonance are also thought 
to be universally connected \cite{braaten_physrep,dincao_jpb}.  
However, when we compare features across the Feshbach resonance, 
we find a systematic discrepancy with theory of a factor of 2 (Table~2). 
This discrepancy can be expressed as a 
difference in the three-body short-range phase 
between the two sides of the 
Feshbach resonance $\Delta\Phi = s_0\, ln (|a^-|/a^+)$ \cite{NielsenMacek,dincao_jpb}.
The locations of the features reported here result in 
phase differences of $0.92(10)(0)$ and $0.86(4)(17)$ 
(the uncertainties are defined in Table~1) 
for the first and second trimer, respectively,
whereas the universal prediction is $1.61(3)$ \cite{braaten_physrep}.
One of the effects of finite temperature is to both broaden the trimer resonances 
and to push them towards smaller $|a|$ \cite{Kraemer,dincao:123201,dincao_jpb}.
This would decrease the values of $\Delta\Phi$ since we extract $L_3$ from a thermal cloud 
at $a^-$ and a much colder BEC at $a^+$.
Measurements of $^{39}$K also show a 
discrepancy with theory across the resonance, but with $\Delta\Phi = 1.91(7)$ \cite{Zaccanti}.
On the other hand, measurements of the first trimer resonance 
and second trimer recombination minimum
in the $|1,0\rangle$ state of $^7$Li result in $\Delta\Phi = 1.7(2)$ 
in good agreement with universal theory, 
assuming the universal scaling of 22.7 between trimer states \cite{Khaykovich}.
These variations in $\Delta\Phi$ may indicate the need for 
additional physics to be included in the universal model \cite{dincao_jpb,henk}.

\bibliographystyle{Science}

\begin{scilastnote} 
\item We thank Evan Olson for his contributions to
this project and acknowledge useful discussions with 
E.~Braaten, 
    J.~P.~D'Incao, 
    C.~H.~Greene, 
    N.~P. Mehta, and H.~T.~C.~Stoof.
Support for this work
was provided by the National Science Foundation, Office of Naval Research,
the Keck Foundation, and the Welch Foundation (C-1133).  
\end{scilastnote}

\noindent {\bf Supporting Online Material}\\
www.sciencemag.org\\
Materials and Methods\\
Figs.~S1, S2, S3, S4\\
References


\clearpage

\parbox[h]{.55\textwidth}{ \begin{tabular}{ll}
\multicolumn{1}{c}{\rule{0pt}{2.6ex} \rule[-1.2ex]{0pt}{0pt} $a>0$} & \multicolumn{1}{c}{$a<0$} \\
\hline
\rule{0pt}{2.6ex} 
    $a^+_1 = 119(11)(0)$             & $a^-_1 = -298(10)(1)$   \\[3pt]
    $a^+_2 = 2676(67)(128)$            & $a^-_2 = -6301(264)(740)$ \\[3pt]
    $a^*_2\;= 608(11)(7)$            & $a^T_{1,1} \sim -120(20)(0)$ \\[3pt]
    $\left[ a^*_{2,1} \approx 1470(15)(38) \right]$     
				& $a^T_{1,2} \approx -295(35)(1)$ \\[3pt]
    $\left[ a^*_{2,2} \approx 3910(60)(278) \right]$      
				& $a^T_{2,1} \approx -2950(200)(150)$ \\[3pt]
    $\eta^+_1 = 0.079(32)(20)$             
				& $a^T_{2,2} \approx -6150(800)(700)$ \\[3pt]
\rule[-1.2ex]{0pt}{0pt} 
    $\eta^+_2 = 0.039(4)(10)$              
				& $\eta^- = 0.13(1)(3)$ \\[3pt]	 
\hline \end{tabular}}
\parbox[h]{.45\textwidth}{\renewcommand{\baselinestretch}{1.1}\normalsize
\noindent {\bf Table~1.}
Locations (in $a_0$) of three- and four-body loss features and 
inelasticity parameters (dimensionless) \cite{som}.
The features $a^*_{2,1}$ and $a^*_{2,2}$ are tentatively assigned.
The first number in parentheses characterizes the range
over which $\chi^2$ of the fit to theory increases by one,
while simultaneously adjusting the other parameters in the fit.
The second number characterizes the systematic uncertainties 
in the determination of~$a$~\cite{som}.
}
\vspace{6\baselineskip}

\parbox[h]{.65\textwidth}{
\begin{tabular}{llcccl}
\rule{0pt}{2.6ex}\rule[-1.2ex]{0pt}{0pt} & Ratio & Data & Theory & $\Delta(\%)$ & \\
\hline
$a > 0$\rule{0pt}{2.6ex} 
	& $a^+_2 / a^+_1$   & $22.5(22)(11)$		& $22.7^*$     & $-1(9)(5)$ &\\[3pt]
	& $a^+_2 / a^*_2$   & $4.40(14)(16)$		& $4.46^*$     & $-1(3)(4)$ &\\[8pt]
	& $a^*_{2,1}/a^*_2$ & $\approx2.42(5)(4)$	& $2.37^\ddag$ & $+2(2)(2)$ &\\[3pt]
	& $a^*_{2,2}/a^*_2$ & $\approx6.4(2)(4)$	& $6.6^\ddag$  & $-3(2)(6)$ &\\
 \\
$a < 0$ & $a^-_2 / a^-_1$   & $21.1(11)(24)$		& $22.7^*$    & $-7(5)(11)$ &\\[8pt]
	& $a^T_{1,1}/a^-_1$ & $\sim$$0.40(7)(0)$	& $0.43^\dag$ & $-6(16)(0)$ &\\[3pt]
	& $a^T_{1,2}/a^-_1$ & $\approx0.99(12)(0)$	& $0.90^\dag$ & $+10(14)(0)$& \\[3pt]
	& $a^T_{2,1}/a^-_2$ & $\approx0.47(4)(4)$	& $0.43^\dag$ & $+9(9)(9)$& \\[3pt]
	& $a^T_{2,2}/a^-_2$ & $\approx0.98(13)(1)$	& $0.90^\dag$ & $+8(14)(1)$& \\[3pt]
  \\
$a \rightarrow \pm\infty $ 
	& $|a^-_1| / a^+_1$ & $2.5(2)(0) $	&  $4.9^*$ & $-49(5)(0)$& \\[3pt]
	& $|a^-_2| / a^+_2$ & $2.4(1)(4)$	&  $4.9^*$ & $-52(2)(9)$& \\[8pt]
	& $|a^-_1| / a^*_2$ & $0.49(2)(1)$	& $0.97^*$ & $-49(2)(1)$& \\[3pt]
	& $|a^-_2| / a^*_2$ & $10.4(5)(14)$	& $22.0^*$ & $-53(2)(6)$& \\[3pt]
\hline
\multicolumn{6}{l}{ {\footnotesize References:
    $^*$ \cite{braaten_physrep}; 
    $^\dag$ \cite{javier};
    $^\ddag$ \cite{dincao:033004}.}
}
\end{tabular}}
\parbox[h]{.35\textwidth}{\renewcommand{\baselinestretch}{1.1}\normalsize
\noindent {\bf Table~2.}
Relative locations of loss features, those predicted by theory,
and the percent difference $\Delta = (\mathrm{Data/Theory} - 1)$.
The uncertainties are those propagated from Table~1.
}

\clearpage\enlargethispage{\baselineskip}
\begin{center}
\includegraphics[height=\textwidth,angle=-90]{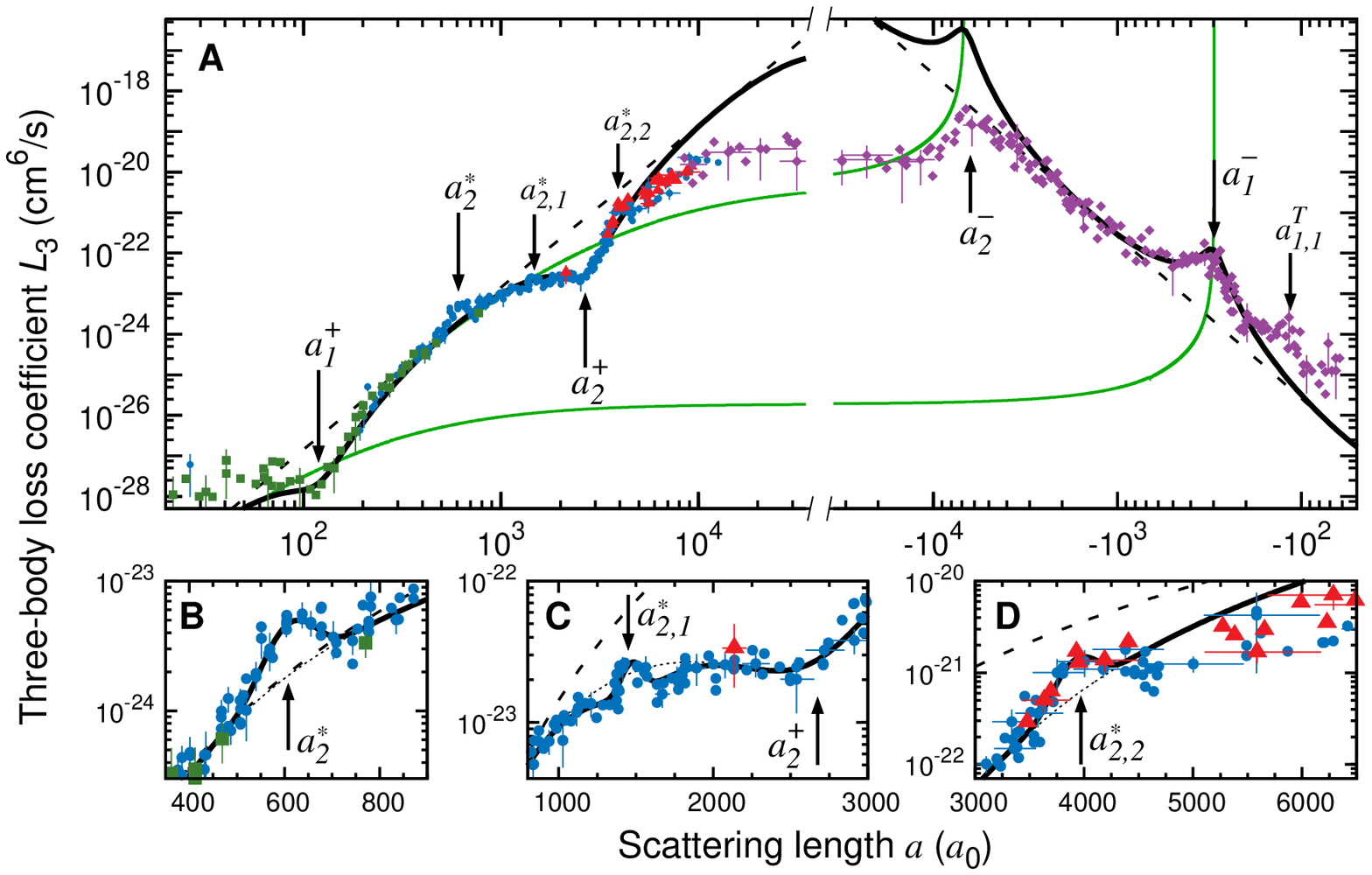}
\end{center}
\noindent {\bf Fig.~1.} 
\renewcommand{\baselinestretch}{1.1}\normalsize
(\textbf{A}) $L_3$ as a function of $a$.
Data shown with ({\color{Purple} $\blacklozenge$})
correspond to a thermal gas with $N \sim 10^6$, $T \sim 1$--$3\,\mu\mathrm{K}$ \cite{theating}, and $U \sim 6\,\mu\mathrm{K}$
and were taken with radial and axial trapping frequencies 
$\omega_r = (2\pi)\,820\,\mathrm{Hz}$ and $\omega_z = (2\pi)\,7.3\,\mathrm{Hz}$, respectively.
The remaining data correspond to a BEC with $N \sim 4 \times 10^5$, $T < 0.5\,T_C$, $U \sim 0.5\,\mu\mathrm{K}$, and $\omega_r = (2\pi)\,236\,\mathrm{Hz}$.
We adjust $\omega_z$ \cite{som} to enhance or reduce three-body loss,
where $\omega_z = (2\pi)\,1.6$\,Hz ({\color{Red} $\blacktriangle$}),
$\omega_z = (2\pi)\,4.6$\,Hz ({\color{RoyalBlue} \large$\bullet$}),
and $\omega_z = (2\pi)\,16\,$Hz ({\color{OliveGreen} $\blacksquare$}).
The dashed lines show an $a^4$ scaling.
The solid thick lines are fits to 
an analytic theory \cite{som,braaten_physrep}.
The thin green lines show the square of the energies, in arbitrary units, 
of the first and second Efimov states
as predicted from the universal theory \cite{braaten_physrep} where we have fixed 
the location of the first Efimov state to overlap with $a^-_1$, 
and the atom-dimer continuum is coincident with the dashed line for $a>0$.
Several representative error bars are shown \cite{som}.
(\textbf{B}---\textbf{D}) Detail around the loss features associated 
with the atom-dimer and two possible dimer-dimer resonances.  
The dotted lines are the fit to Eq.~S4,
while the solid lines include additional superimposed Gaussian fits
to account for the features not described by Eq.~S4.

\begin{center}
\includegraphics[height=.8\textwidth,angle=-90]{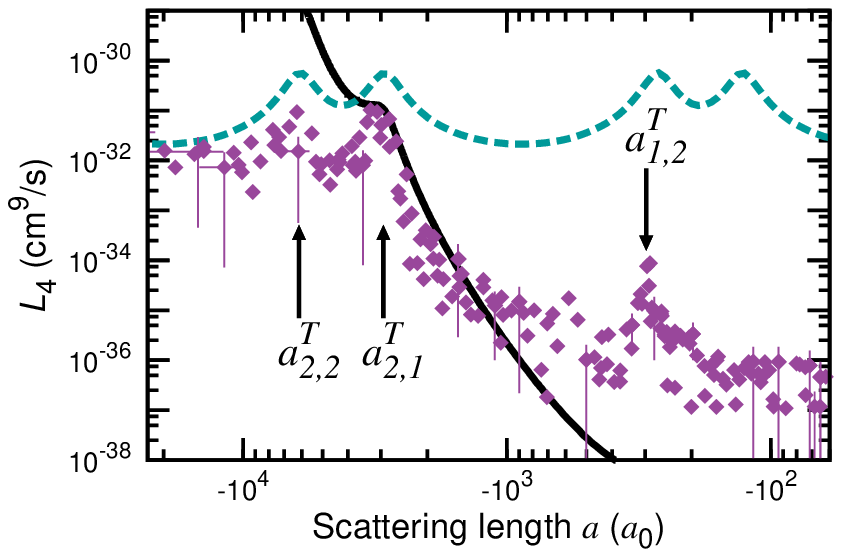}
\end{center}
\noindent {\bf Fig.~2.} 
\renewcommand{\baselinestretch}{1.1}\normalsize
$L_4$ extracted from a thermal gas.
The solid curve is motivated by theory \cite{som,mehta},
and the dashed curve is the solid curve divided by $a^7$ \cite{wang:133201}.
The uncertainty in $L_4$ from the fit is a factor of 2,
while the sytematic uncertainty is
a factor of 3 due to uncertainties
in $\omega_r$, $\omega_z$, $N$ and $T$.
For $|a| > 2 \times 10^4\,a_0$ differentiation between three- and four-body
losses becomes unreliable due to the very fast decay rates.
Data with $L_4 < 10^{-36}\,\mathrm{cm^9/s}$ are consistent with no four-body loss.

\clearpage
\renewcommand{\theequation}{\mbox{S\arabic{equation}}}

\noindent {\large Supporting Online Material for}\\[6pt]
{\Large\bf Universality in Three- and Four-Body Bound States of Ultracold Atoms}\\[6pt]
\noindent S. E. Pollack, D. Dries, R. G. Hulet

\section*{Materials and Methods}

A set of non-Helmholtz coils are used to add or subtract additional axial confinement 
in the hybrid magnetic plus optical dipole trap used in the experiment.
The radial trapping frequency $\omega_r$ is determined from atom loss by parametric excitation, 
and the axial trapping frequency $\omega_z$ is determined from collective dipole oscillations.

\subsection*{Determination of Scattering Length}

The $s$-wave scattering length $a$ is controlled via a magnetic Feshbach resonance \cite{s_frucg}.
We extract $a$ (for $a > 0$) as
a function of magnetic field $B$ from the axial size of a Bose-Einstein
condensate \cite{s_Pollack}.  
The measured functional form of $a$~vs.~$B$ is well described
by a Feshbach resonance fit $a (B) = a_{BG} [ 1 + \Delta/(B - B_\infty)]$,
where the values $a_{BG} = -24.5^{+3.0}_{-0.2}\,a_0$, $\Delta = 192.3(3)\,$G,
and $B_\infty = 736.8(2)\,$G were previously reported \cite{s_Pollack}.
The standard deviation of the residuals from the Feshbach resonance
fit is 15\% for $a < 10^3\,a_0$ and 30\% for $a > 10^3\,a_0$ (Fig.~S2).

To repeatably achieve very large values of $a$ it is necessary to have 
both high field stability and accurate knowledge of the location of $B_\infty$.
We determine the shot-to-shot stability and calibration of the magnetic field from
radio frequency spectroscopy on the $|1,1\rangle \rightarrow |2,2\rangle$ transition.
We have improved the control of the current in the coils that provide the magnetic bias field
in our experiment such that a Lorentzian characterizing the shot-to-shot 
field stability has a full width at half maximum of 115\,kHz, 
corresponding to 42\,mG at a bias field of 717\,G (Fig.~S3C).
With this improved field stability we have 
increased the precision in the determination of the resonance location to
$B_\infty = 736.97(7)\,$G.
The uncertainty in $B_\infty$ is dominated by systematic uncertainty 
in the extracted values of $a$ from the measured axial sizes \cite{s_Pollack}.
The fractional uncertainty in the determination of $a$ is given by
$\delta a/a = \delta B / (B - B_\infty) \approx 1.5 \times 10^{-5}\,a/a_0$,
where $\delta B$ is dominated by the uncertainty in $B_\infty$.

Since we have only measured $a$ for $a>0$, we have no
direct knowledge of $a<0$.  
However, a coupled-channels calculation \cite{s_Houbiers} agrees
with the Feshbach resonance fit to within 10\% over the range
of $10 < a/a_0 < 4 \times 10^4$ (Fig.~S3) which gives us confidence that
the Feshbach resonance fit is equally reliable on the $a<0$ side of the resonance.

\subsection*{Determination of the Loss Coefficients}

Extraction of $L_3$ and $L_4$ from the measured atom number loss curves 
$N(t)$ requires the evaluation of the spatially-averaged moments of the density distribution 
$\langle n^2 \rangle$ and $\langle n^3 \rangle$.
By comparing the measured distributions with 
a Thomas-Fermi inverted parabola in the case of a pure Bose-Einstein condensate,
we find to a good approximation
that the distributions remain in thermal equilibrium throughout the decay process.
For a condensate, the axial Thomas-Fermi radius is
$R = (15 \hbar^2 \omega_r^2 N a / m^2 \omega_z^4)^{1/5}$,
the peak density is $n_0 = (15 N \omega_r^2)/(8\pi R^3 \omega_z^2)$,
and $\langle n^2 \rangle = \gamma^{2/5} N^{4/5}$,
where $\gamma = (25\,m^{6} \omega_r^4 \omega_z^2) / (6272 \sqrt{42}\,\pi^{5} \hbar^{6} a^3)$.
The observed decay fits well to a purely three-body loss process for a condensate,
so we neglect $L_4$ in this case.  
Since we are not explicitly fitting for $L_4$,
four-body effects if present may lead to an increase in the 
extracted loss rate $L_3$ \cite{s_javier}.
The decay is then described by
\begin{equation}
\frac{1}{N}\frac{dN}{dt} = - \frac{g^{(3)}}{3!} L_3 \gamma^{2/5} N^{4/5},
\end{equation}
which has the solution
\begin{equation}
N(t) = \frac{N_0}{\left(1 + \displaystyle \frac{ 4}{5} 
	\frac{g^{(3)} L_3}{3!} \gamma^{2/5} N_0^{4/5} 
	t \right)^{5/4} }.
\end{equation}

A thermal gas is well described by a cylindrically-symmetric Gaussian where
$\langle n^2 \rangle = n_p^2 / \sqrt{27}$,
$\langle n^3 \rangle = n_p^3 / 8$,
and the peak density is $n_p = N (\omega_z/ \omega_r) [m \omega_r^2 / 2 \pi k_B T]^{3/2}$.
Heating due to recombination is expected to become important when $\epsilon \lesssim U$ \cite{s_weber}.
However, there is no appreciable change observed in the Gaussian width during the decay
even though the loss mechanism preferentially targets atoms at higher densities.
This may be due to a lack of rethermalization during the decay \cite{s_ferlaino:140401}.
We find that both $L_3$ and $L_4$ contribute to the loss for the thermal gas.  Since
we have not found a closed-form solution to Eq.~1, we instead use the following
implicit solution to extract $L_3$ and $L_4$:
\begin{eqnarray}
t = \frac{3\sqrt{3}}{2 n_p^2 L_3} \left[\left(\frac{N_0}{N}\right)^2-1 \right]
	   + \frac{27 L_4}{8 n_p L_3^2} \left(1-\frac{N_0}{N}\right) 
	 - \frac{81 \sqrt{3} L_4^2}{64 L_3^3} \log \left[ \left(\frac{N}{N_0}\right) 
    \frac{8\sqrt{3}L_3 + 9 L_4 n_p} {8\sqrt{3}L_3 + 9 L_4 n_p (N_0/N)} \right],
\end{eqnarray} 
where we have assumed $g^{(3)} = 3!$ and $g^{(4)} = 4!$ for a non-condensed gas.

In Fig.~1 the vertical error bars correspond to the range in $L_3$ for which 
the $\chi^2$ of the fit to Eq.~S3 increases by one, 
while simultaneously adjusting $L_4$ and $N_0$ to minimize $\chi^2$.
Systematic uncertainties in $\omega_r$, $\omega_z$, $N$, and $T$,
which are not included in these error bars, 
contribute as much as a factor of $2$ in the uncertainty of $L_3$.
The representative horizontal error bars are due to shot-to-shot variation in the
magnetic field and the determination of $a$ from the Feshbach resonance fit.
Background loss limits the sensitivity of the measurement to 
$L_3 > 2(1) \times 10^{-28}\,\mathrm{cm^6/s}$.
The error bars in Fig.~2 are similarly determined.

\clearpage
\subsection*{Comparing with Theory}

The universal theory \cite{s_braaten_physrep} describing Efimov physics
predicts that the three-body loss rate coefficient 
is described by $L_3(a) = 3 C(a) \hbar a^4 / m$ where
$C(a)$ is a logarithmically periodic modulation.  
The following expression describes this modulation:
\begin{equation}
C(a) = \left\{ 
    \begin{array}{l@{\quad\quad}l}
	\displaystyle \frac{4590\, \sinh(2 \eta^-)}{ \sin^2\left(s_0 \ln(a/a^-) \right) 
	    + \sinh^2\eta^- } 				& (a<0), \\ \\
	67.12 e^{-2 \eta^+} \left[ \sin^2\left(s_0 \ln (a/a^+) \right) 
	    + \sinh^2 \eta^+ \right] + 16.84 (1 - e^{-4 \eta^+}) 	& (a>0),
    \end{array} \right. 
\end{equation}
where the first and second terms for $a>0$ account for coupling to weakly- 
and deeply-bound dimer states, respectively \cite{s_braaten_physrep,s_PhysRevLett.83.1751}.
The value $a^-$ denotes the resonance location when the energy of the Efimov trimer is degenerate
with the free atom continuum, and the value $a^+$ is the location
of a recombination minimum \cite{s_NielsenMacek}.
This expression is log-periodic with $C( e^{\pi/s_0} a ) = C(a)$,
where the universal parameter $s_0 = 1.00624$ is known 
from theory \cite{s_braaten_physrep,s_efimov71}.

The four-body loss coefficient $L_4$ is predicted to have a 
similar form to that of $L_3$:
\begin{equation}
L_4 (a,a^T) = 4\,C_4\,\frac{\hbar |a|^7}{m} 
	\frac{\sinh(2\eta^-)}{\sin^2\left(s_0 \ln(a/a^T) \right) + \sinh^2\eta^- } 
	    \quad\quad (a<0),
\end{equation}
where $C_4$ is a theoretically undetermined universal constant \cite{s_mehta}. 
Eq.~S5 is phenomenologically derived from the theory of Ref.~{S11} \cite{s_greenePC}.
We find that $C_4 = 16(8) \times 10^4$ in the region $1000 < -a/a_0 < 2500$, assuming 
that $\eta^- = 0.13$, as for the three-body resonance.
In Fig.~2 we plot 
$\frac{1}{2}\{ L_4(a, 0.90\,a^-_{1}) + L_4(a, 0.43\,a^-_{1})\}$
where we have replaced $a^T$ with 
the predicted locations of the two tetramer states linked to the
first trimer state \cite{s_javier,s_mehta}.\\

\clearpage
\begin{center}
\includegraphics[height=\textwidth,angle=-90]{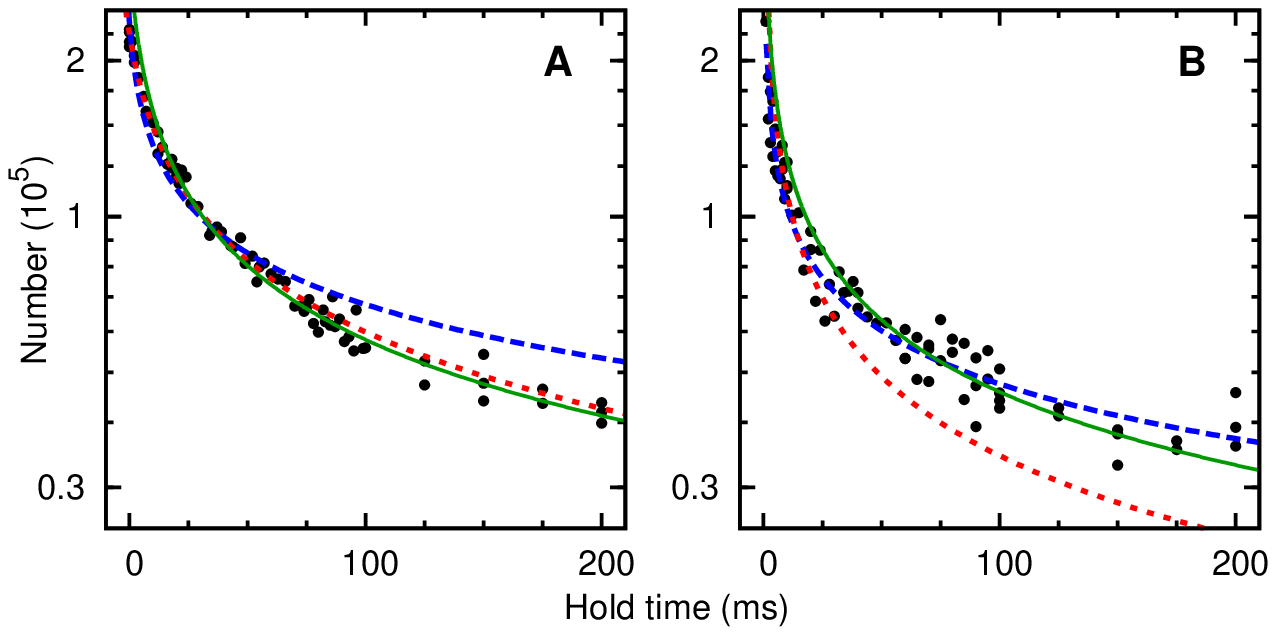}
\end{center}
\noindent {\bf Fig.~S1.}
Loss dynamics at two values of $a < 0$ for a thermal gas.
The dots are data.
The dotted red line is a fit of the data to the solution of Eq.~1
with only three-body loss accounted for, the dashed blue line is the fit when
only four-body loss is included,
and the solid green line is a fit accounting for both effects (Eq.~S3).
(\textbf{A}) $a = -1800\,a_0$, where three-body losses dominate;
(\textbf{B})~$a = -3300\,a_0$, near $a^T_{2,1}$ where four-body losses dominate.

\begin{center}
\includegraphics[height=\textwidth,angle=-90]{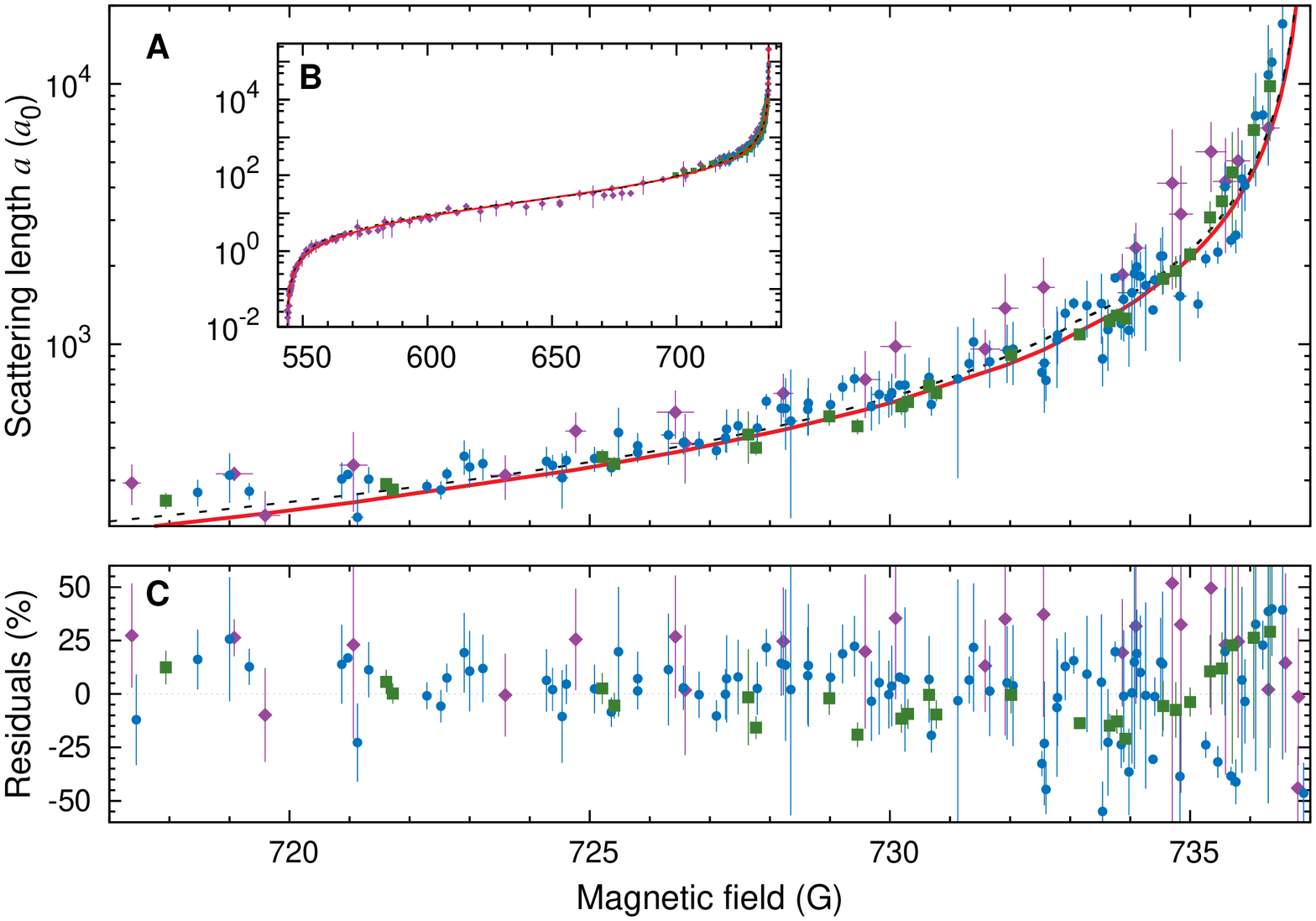}
\end{center}
\noindent {\bf Fig.~S2.} (\textbf{A}) $a$ extracted from the axial
size of Bose-Einstein condensates
as a function of magnetic field.
Results of a coupled-channels calculation are shown by the solid red line. 
The dashed black line is the Feshbach resonance fit. 
({\color{Purple}  \large$\blacklozenge$}) 
Data previously reported 
with trapping frequencies $\omega_r = (2\pi)\,193\,\mathrm{Hz}$ and $\omega_z = (2\pi)\,3\,\mathrm{Hz}$ \cite{s_Pollack}.
Data with $\omega_r = (2\pi)\,236\,\mathrm{Hz}$ and 
$\omega_z = (2\pi)\,4.6\,\mathrm{Hz}$ ({\color{RoyalBlue} $\bullet$}) or 
$\omega_z = (2\pi)\,16\,\mathrm{Hz}$ ({\color{OliveGreen} $\blacksquare$}).
Beyond mean field effects become important when $n_0 a^3 \gtrsim 0.1$ \cite{s_LHY}.
We apply a mean field correction for data with $0.1 < n_0 a^3 < 1$,
and omit data with $n_0 a^3 > 1$ in the Feshbach resonance fit \cite{s_Pollack}.
(\textbf{B}) Full range of data spanning 7 decades in~$a$.
(\textbf{C}) Fractional residuals of the extracted values of $a$ from
the Feshbach resonance~fit.

\begin{center}
\includegraphics[height=\textwidth,angle=-90]{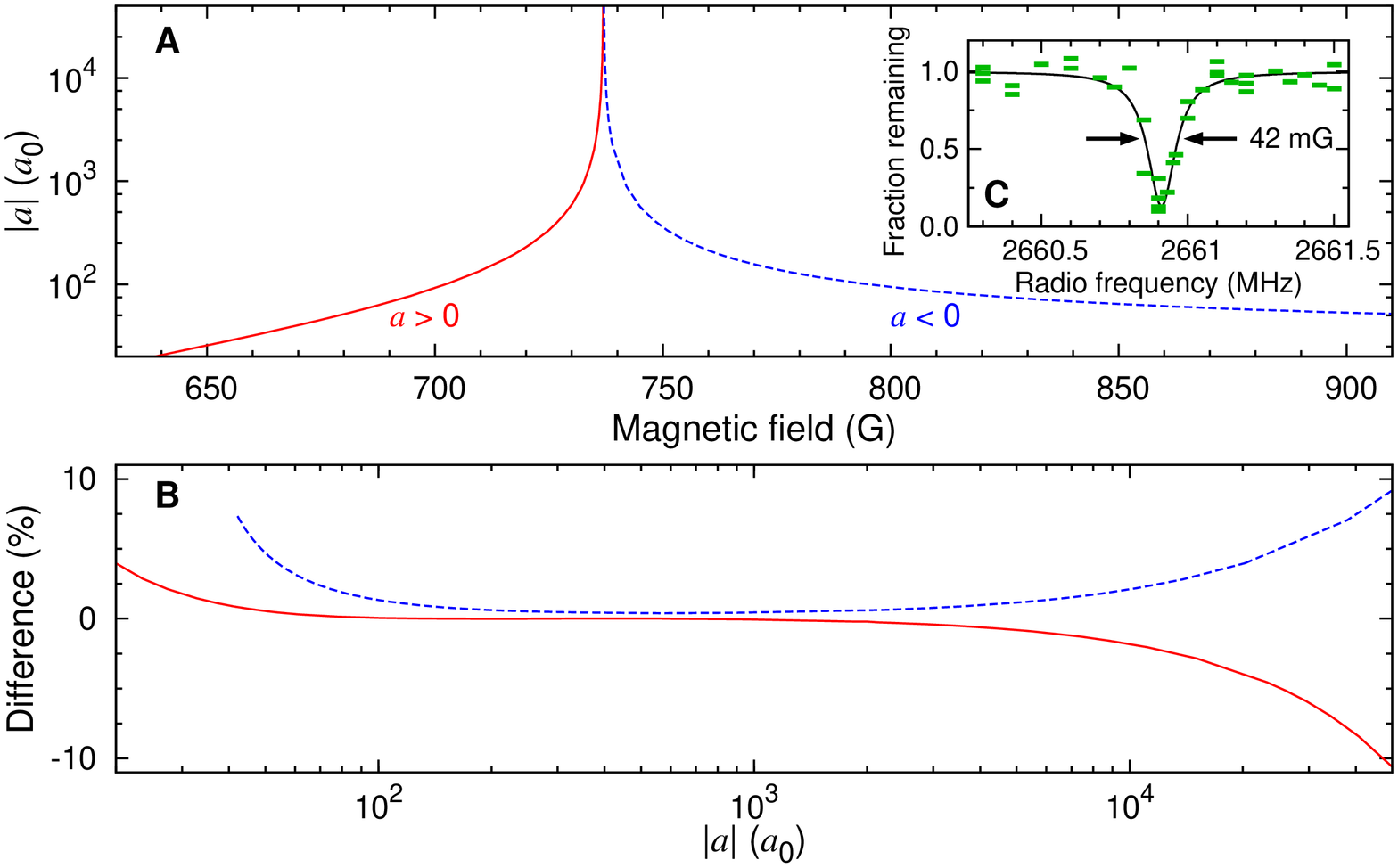}
\end{center}
\noindent {\bf Fig.~S3.} (\textbf{A}) $a$ vs. magnetic field 
from a coupled-channels calculation.  
(\textbf{B}) Fractional difference between the coupled-channels calculation
and the Feshbach resonance fit used to determine~$a$
(solid red line $a > 0$, dashed blue line $a < 0$).
(\textbf{C}) Radio frequency spectroscopy signal at 717\,G
showing a full width at half maximum of~115\,kHz.

\begin{center}
\includegraphics[height=\textwidth,angle=-90]{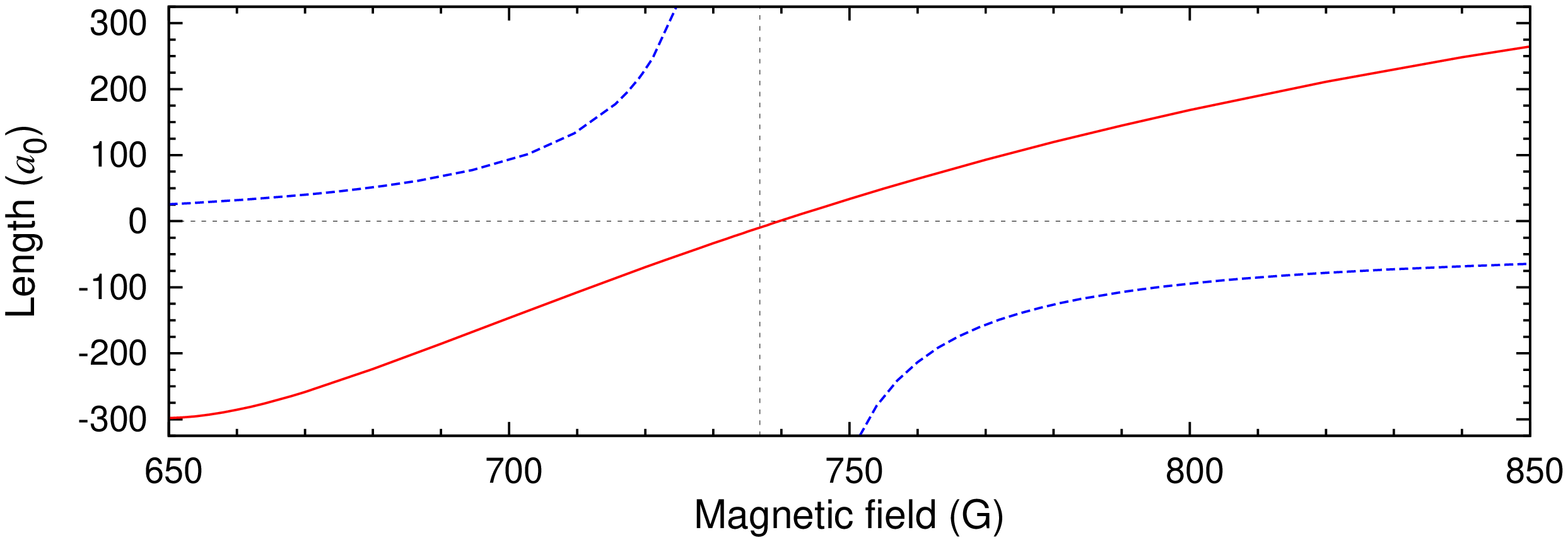}
\end{center}
\noindent {\bf Fig.~S4.} The effective range $R_e$ (solid red) and
scattering length $a$ (dashed blue) vs. magnetic field,
extracted from a coupled-channels calculation through
a low energy expansion $k \cot \delta = - 1/a + R_e k^2/2$, where
$\delta$ is the scattering phase shift \cite{s_frucg}.
The dotted vertical line is the location of $B_\infty$.



\begin{thebibliography}{10}
\bibitem{som}
Materials and methods are available as supporting material on Science Online.
\bibitem{theating}
In ramping from $-200\,a_0$ to $a < -3000\,a_0$ we observe an increase in the
  axial size of the thermal cloud which is consistent with a temperature
  increase of the cloud to about $3\,\mu\mathrm{K}$. During the trap loss
  measurements we observe negligible change in the Gaussian width of the
  thermal cloud \cite{som}.
\end{thebibliography}
\end{document}